# EOF Library: Open-Source Elmer and OpenFOAM Coupler for Simulation of MHD With Free Surface

J. Vencels , A. Jakovics, V. Geza and M. Scepanskis

**Abstract**

In this work, we present an open-source library called EOF which provides efficient two-way coupling between the finite element multi-physics simulation software Elmer and computational fluid dynamics software OpenFOAM. Library controls interpolation and communication of fields between two codes. The main focus of this code is industrial magnetohydrodynamics with free surface, but it can be extended to other 2D and 3D multi-physics problems. We present simulation results for 2D axisymmetric problem of conductive fluid with free surface surrounded by alternate electromagnetic field created by inductor and compare them with results of analogous simulation.

**Introduction**

Manipulation of conductive fluids with electromagnetic field is of great importance for industrial technologies, such as growth of semiconductor crystals, liquid-metal pumps for cooling nuclear reactors and metallurgical casting. Manufacturing optimization, improvement of material quality and development of new technologies often relies on coupled physical processes like electromagnetics, fluid dynamics, multiphase flows and heat transfer.

From spatial and temporal discretization viewpoint the great difference in scales involved in turbulent fluid flows, high electromagnetic frequencies and skin effect makes numerical studies of magnetohydrodynamics (MHD) problems computationally challenging. While commercial codes provide multi-physics and parallelization capabilities, they have well known limitation of being closed black boxes. There is a high demand for open tools that could be adopted for specific problems and run on supercomputers. Considering that open-source science-oriented MHD codes for astrophysical plasmas are common, the same is not true for industrial multi-physics applications.

In this paper, we present our open-source library called EOF which provides two-way coupling between Elmer [1] - finite element method (FEM) code and OpenFOAM [2] - finite volume method (FVM) code. EOF is collection of add-on libraries that are dynamically linked to Elmer and OpenFOAM, it is written in Fortran and C++, and uses the same style as original codes. EOF takes care of interpolation and communication of fields and leaves physics development in hands of user. Library can communicate any variables between two codes, consequently users also could find it useful for applications other than MHD.

## 1. Physical background

When time harmonic magnetic field is applied on conductive fluid then electric current is induced. This current interacts with magnetic field and produces time averaged Lorentz force $F_L$:

$$\vec{F_L} = \frac{1}{2} Re(\vec{J} \times \vec{B}^*), \qquad (1)$$

where B - magnetic flux density, J - current density and operator "*" is complex conjugate.

Time averaged Lorentz force is applied on incompressible Navier-Stokes equation which is momentum equation of fluid:

$$\frac{\partial \vec{U}}{\partial t} + (\vec{U} \cdot \nabla)\vec{U} - \nu \nabla^2 \vec{U} = \vec{F_g} + \vec{F_L}, \qquad (2)$$

where U - velocity, $\nu$ - kinematic viscosity, $F_g$ - gravitational force.

## 2. Simulation Software

We identified five main requirements for industrial MHD simulation software:

- Fluid and electromagnetic solvers capable solving transient 3D problems
- Complex notation for time-harmonic fields in electromagnetic solver
- Volume of Fluid (VOF) free surface and high Reynolds turbulence models
- Integration with pre- and post-processing tools
- Handling complex geometries

To our knowledge no-one open-source code can meet all these requirements. FVM is mainly used for computational fluid dynamics (CFD) due its robustness and conservation properties. FEM is common for electromagnetic (EM) solvers since higher order basis functions result in better accuracy [3]. In our case coupling two codes based on FVM and FEM is natural choice.

### 2.1. Elmer

Elmer [1] is multiphysics simulation software based on FEM and written in Fortran. The main applications include structural mechanics, ice sheet modelling and electromagnetics. While Elmer aso has CFD solver, we found it incapable to solve flows in highly turbulent regime.

### 2.2. OpenFOAM

OpenFOAM [2] is leading open-source software for CFD, based on FVM and written in C++. It has electrostatic, magnetic and mhd solvers, but it lacks complex variable notation for time-harmonic fields and does not have proper boundary conditions for magnetic fields between multi-regions. It is unsuitable for EM simulations with complex geometries.

## 3. Two-way Coupling

In this section we describe how two-way coupling for conductive fluid with free surface surrounded by alternate EM field is implemented. See schematic of the problem in Fig. 1 (a).

Both codes are running simultaneously. Computation is performed by one code, the other code at that moment is idle and waiting. Typically EM domain is larger in size than fluid domain since it includes coils, vessels, etc. Fluid solver's domain includes regions where air and melt mixture is present, in most cases it has one region.

Computation scheme is shown in Fig. 1 (b). OpenFOAM solves fluid with free surface, computes electrical conductivity of air and melt mixture

$$\sigma_{mix} = \alpha \cdot \sigma_{melt}, \tag{3}$$

where $\alpha \in [0,1]$ and $\sigma_{melt}$ are melt's fraction and conductivity respectively, and sends it to Elmer. Elmer solves EM problem, computes Lorentz force (1) and sends it to OpenFOAM. This cycle is performed until the last time step.

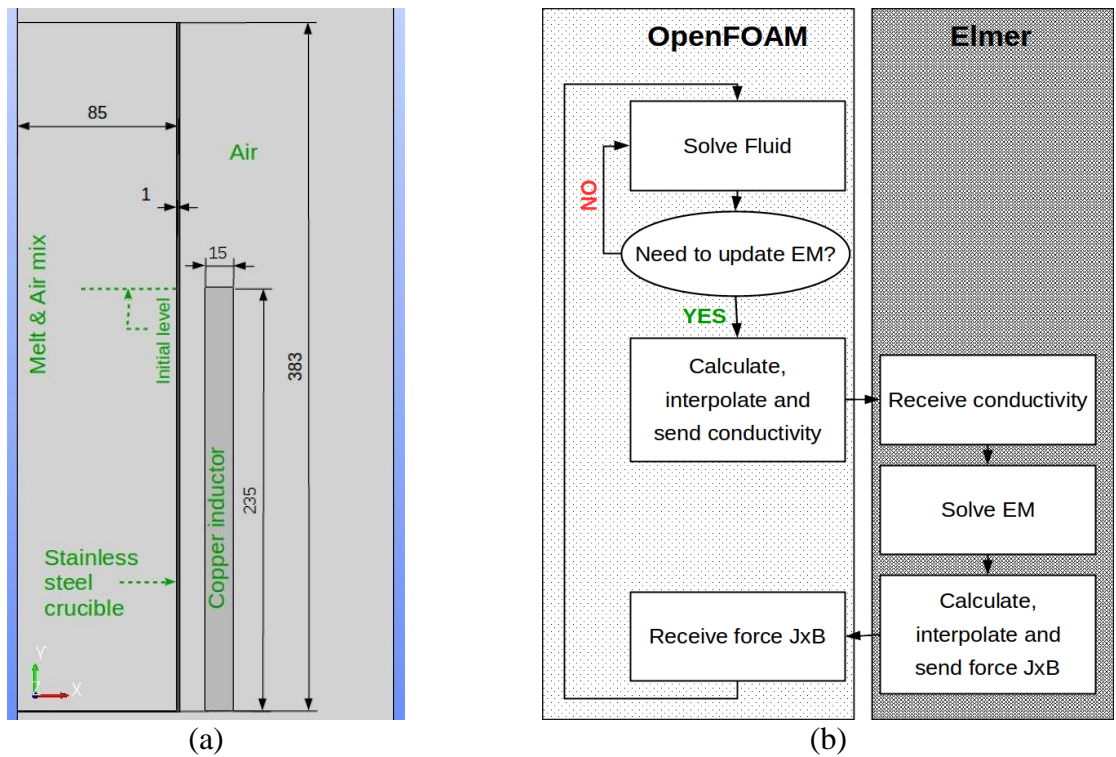

Fig. 1. Schematic drawing for 2D axisymmetric problem (a) and computational scheme (b)

### 3.1. Field Communication with Message Passing Interface

File-based coupling may look as a straightforward solution for field communication between two stand-alone codes, but it is barely efficient and a better way exists. Both OpenFOAM and Elmer are using Message Passing Interface (MPI) [4] for communication between parallel processes. Without breaking internal parallelization we can use MPI for communication between any number of simultaneously running programs.

For example, to run two MPI-enabled programs simultaneously we execute the following command in terminal

$ mpirun -n [nCores1] [program1] : -n [nCores2] [program2]

where [nCores1] and [nCores2] are the number of processes used for programs [program1] and [program2] respectively.

*3.2. Field Interpolation Between Meshes*

Solving EM and fluid problems on the same mesh is impractical. EM domain usually is larger than fluid domain since it includes inductors, air regions, etc. In addition, resolving skin depth in EM or highly turbulent flows require finer mesh. Efficient and accurate interpolation between parallelly distributed, non-conforming meshes is very important. For interpolation procedure the most time consuming part is searching corresponding cell centres and nodes, for that reason we do this procedure only once during the initialization phase.

For Elmer to OpenFOAM interpolation we use Elmer's built-in function *InterpolateMeshToMeshQ* which creates projector matrix from one mesh to another. During simulation, interpolation is repeatedly done by multiplying projector matrix and vector containing variable values. This is fast operation and interpolation has accuracy comparable to the order of used elemental basis functions (at least 1st order).

During initialization phase, OpenFOAM to Elmer interpolator searches and saves found element indexes to array. When simulation is run, interpolation is repeatedly done by taking values at cell centers. This is fast operation, but with 0th order accuracy.

*3.3. Adaptive Criteria for Updating Electromagnetic Solution*

Fluid dynamics in most cases require smaller time step than EM and in laminar regime take less computational time per time step. Reducing frequency with which EM solver is called can reduce the total simulation time, in some cases more than by half. Such example is steady flow when there is no need to update Lorentz force distribution.

We use condition

$$max(abs(\alpha_{old} - \alpha)) > 0.5, \qquad (4)$$

where α and α_old are melt's fraction at current time step and last time step when the EM solver was called. Both α and α_old can have values between 0 (air) and 1 (fluid). If fluid fraction has changed more than by 0.5 at any cell, then Lorentz force is updated. User can set any other threshold value is needed. With this criteria, Elmer typically is called about once per 5-20 OpenFOAM iterations.

**4. Results**

We solve 2D axisymmetric problem of conductive fluid with free surface which is surrounded by alternate electromagnetic field created by inductor, see Fig. 1 (a). This problem was described and solved by Spitans et al. [5] using externally coupled ANSYS Classic and ANSYS CFX commercial software. Authors verify their results against experimental data.

Melt (Wood's metal) has electrical conductivity $\sigma_{melt} = 10^6 \, S/m$, kinematic viscosity $\nu_{melt} = 4.47 \cdot 10^{-7} \, m^2/s$, density $\rho = 9400 \, kg/m^3$. Although surface tension was not specified, we use $\gamma = 1 \, N/m$ and contact angle $\theta = 90°$. Electric conductivity of stainless steel crucible $\sigma_{cruc} = 1.43 \cdot 10^6 \, S/m$. Inductor has homogeneous current distribution with density $J_{ampl} = 1.185 \cdot 10^7 \, A/m^2$ and frequency $f = 330 \, Hz$. Turbulence model is $k - \varepsilon$, max Courant number set to 0.5 and simulation is run until $t_{end} = 3 \, s$.

Running simulation in parallel on 8 cores (4 for OpenFOAM and 4 for Elmer) takes 5 hours. Elmer mesh has 77k elements, OpenFOAM mesh has 75k cells.

In Fig. 2 (a) Lorentz force density and flow pattern at steady regime is shown. Comparing our results with [5] we found that Lorentz force is in good agreement, while our flow velocity is larger by 30%.

In Fig. 2 (b) we compare obtained melt's shape (gray area) with analogous simulation result from [5] (curved dotted line) and initial filling (horizontal dotted line). The height and shape of meniscus is in good agreement in the middle of crucible, but deviates closer to the wall.

In Fig. 3 shape of the melt is shown at different times.

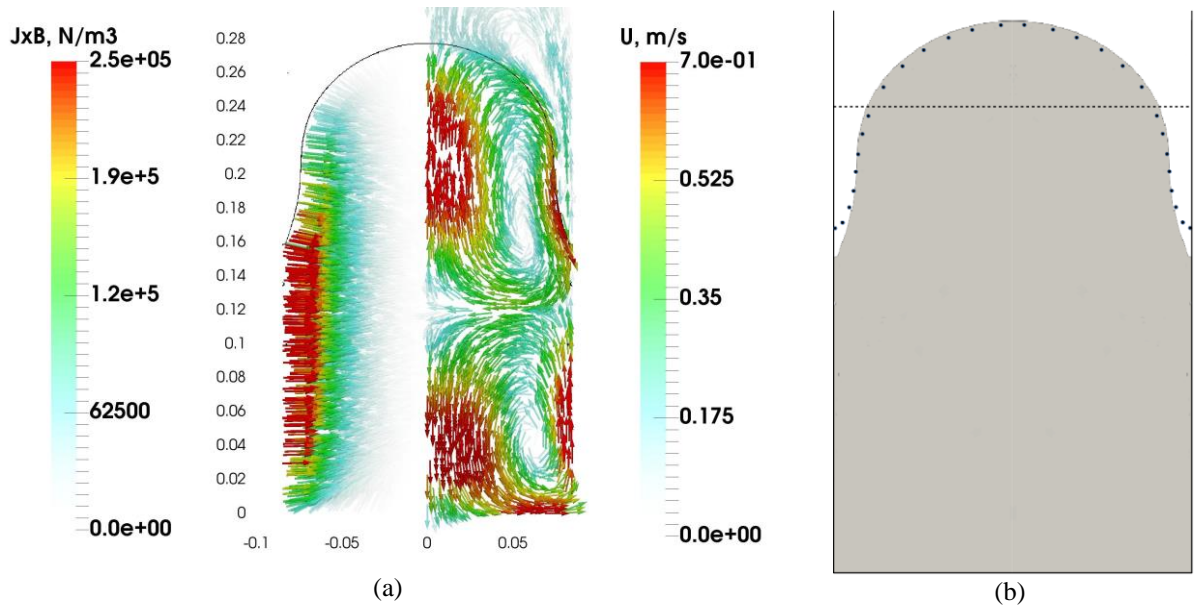

(a)          (b)

Fig. 2. (a) Lorentz force density and flow pattern at steady regime t=3s. (b) shape of melt (gray area), analogous simulation result from [5] (curved dotted line) and initial filling (horizontal dotted line).

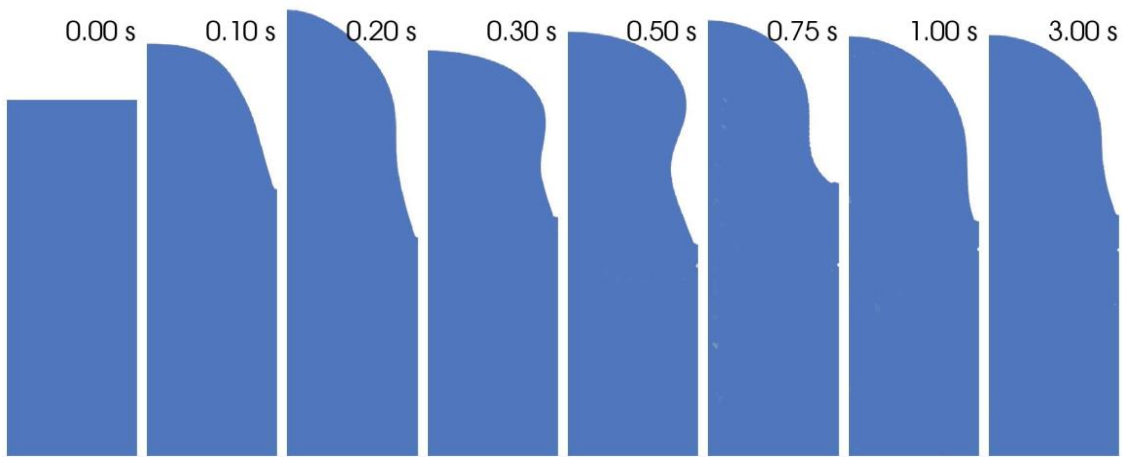

Fig. 3. Shape of the melt at different times

## Conclusions

Open-source library called EOF which provides two-way coupling between Elmer and OpenFOAM was developed. EOF does interpolation and communication of fields between two codes and could be used for wide range of multi-physics applications.

We repeated 2D axisymmetric problem of conductive fluid with free surface and got good agreement for Lorentz force density and meniscus shape. Our flow velocity was larger by 30% which could be related to different turbulence models.

Future work will focus on implementation of heat equation, viscosity and electrical conductivity dependence on temperature.

## Acknowledgments

The authors would like to thank Sergejs Spitans for sharing his valuable experience in coupling electromagnetic and fluid codes, and his guidance in interpreting simulation results.

## Authors

M.Sc. Vencels, Juris  
E-mail: juris.vencels@gmail.com

Dr. Phys. Jakovics, Andris  
E-mail: andris.jakovics@lu.lv

Dr. Phys. Geza, Vadims  
E-mail: vadims.geza@lu.lv

Dr. Phys. Scepanskis, Mihails  
E-mail: mihails.scepanskis@lu.lv

Faculty of Physics and Mathematics  
University of Latvia  
Zellu str. 23  
LV-1002 Riga, Latvia